\documentstyle[11pt]{article}

\begin{document}

\title{SPIN-OTHER-ORBIT OPERATOR IN THE TENSORIAL FORM OF SECOND QUANTIZATION}
\author{Gediminas Gaigalas, Andrius Bernotas and Zenonas Rudzikas \\
{\em State Institute of Theoretical Physics and Astronomy,} \\
{\em A. Go\v{s}tauto 12, 2600 Vilnius, LITHUANIA} \\
\ \\
Charlotte Froese Fischer \\
{\em Department of Computer Science, Box 1679 B,} \\
{\em Vanderbilt University, Nashville, TN 37235, USA}}
\date{}
\maketitle

\vspace{1.5in} {\bf PACS: 3110, 3115, 3130}

\clearpage

\begin{abstract}
The tensorial form of the spin-other-orbit interaction operator in the
formalism of second quantization is presented. Such an expression is needed
to calculate both diagonal and off-diagonal matrix elements according to an
approach, based on a combination of second quantization in the coupled
tensorial form, angular momentum theory in three spaces (orbital, spin and
quasispin), and a generalized graphical technique. One of the basic
features of this approach is the use of tables of standard quantities,
without which the process of
obtaining matrix elements of spin-other-orbit interaction operator between
any electron configurations is much more complicated. Some special cases are
shown for which the tensorial structure of the spin-other-orbit interaction
operator reduces to an unusually simple form.
\end{abstract}

\clearpage

\section{Introduction}

The spin-other-orbit interaction operator is one of the most complex operators
occurring in atomic structure calculations and accounts for the
relativistic corrections in the Breit-Pauli approximation. Because of its
complexity this operator has deserved special attention from a number of
authors, and various modifications of its expression are known from the
literature (\cite{JS} and references therein,~and \cite{HS,GH,ASTC}). In
practical applications the most acceptable modification is the one where the
operator of the spin-other-orbit interaction has the simplest analytical
structure and, at the same time is well formalized to use in the programs
based on methods of atomic structure calculations. From this point of view
the expression derived by Glass and Hibbert \cite{GH} is convenient, and it
is used in functioning computer code MCHF\_ASP \cite{MCHFASP}.
Still, an efficient approach of angular integrations
developed by Gaigalas and Rudzikas \cite{GR} (later on referred to as P1)
and Gaigalas, Rudzikas and Froese Fischer \cite{GRF}
(later on referred to as P2) makes the calculations up to 7 times faster than
calculations based on other methods.

In P1~\cite{GR} the
combination of second quantization in coupled tensorial form, angular
momentum theory in three spaces (orbital, spin and quasispin) and a
generalized graphical technique was proposed to formalize the angular
integrations when up to two open shells are involved. In P2~\cite{GRF}
the formalization was developed further so that
configurations with arbitrary number of open shells are included. This
approach poses new requirements of its own, mainly because of the
use of standard quantities. In particular, we seek the following
goals. First, we need to obtain an expression of the spin-other-orbit
operator in second quantization formalism in order to use the tables of
submatrix elements of standard quantities (analogous to those of $U^k$ and $%
V^{k1}$) while obtaining not just diagonal matrix elements like in Jucys and
Savukynas \cite{JS}, but also off-diagonal matrix elements, namely
off-diagonal with respect to the
configuration's matrix elements. Second, we need a series of explicit formulae
with practical recommendations for their subsequent use according to the
approach described in P1~\cite{GR} and P2~\cite{GRF}, which would allow
us to exploit the quasispin
formalism (Rudzikas and Kaniauskas \cite{RK}) and to take advantage of
having recoupling matrices simpler than those in the approach used by Glass
and Hibbert \cite{GH} in $LSJ$ coupling or Grant \cite{Grant} in $jj$
coupling.

In the second section of this paper we sketch a way of obtaining the general
expression for the spin-other-orbit interaction operator in the coupled
tensorial form of second quantization. In the third section we present
explicit expressions for the submatrix elements occurring in the amplitude
parts of this operator. These values are necessary for the calculation of
matrix elements of the spin-other-orbit interaction operator between
arbitrary configurations (see P2~\cite{GRF}). In
the fourth section we discuss simplifications that are possible in some
special cases of electron distributions in subshells acted upon. This allows
us to reduce the amount of spin-angular integration.

\section{Spin-Other-Orbit Interaction}

   From Eq. (24-18) of~Slater \cite{S} we have the two-particle part of
spin-orbit interaction, also called the spin-other-orbit interaction,
between electrons $i$ and $j$ (using a.u. instead of Rydberg as the unit of
energy) as
\begin{equation}
\label{eq:b}H_{ij}^{soo}=\frac{\alpha ^2}2\left( \left\{ -\frac
1{r_{ij}^3}\left[ {\bf r}_{ij}\times {\bf p}_i\right] +\frac
1{r_{ij}^3}\left[ {\bf r}_{ij}\times 2{\bf p}_j\right] \right\} \cdot {\bf s}%
_i\right) \left( 1+P_{ij}\right),
\end{equation}
where $\alpha =7.29735308\cdot 10^{-3}$ is the fine structure constant in atomic
units, and $P_{ij}$ is the operator of permutation $i\rightleftharpoons j$
of electrons acting upon the expression preceding it.

Definition (\ref{eq:b}), along with the angular momentum theory identity

\begin{equation}
\label{eq:a}{\bf L}_{jk}=\left[ {\bf r}_{jk}\times {\bf p}_j\right]
\end{equation}
gives us


\begin{eqnarray}
\label{eq:bb}
H_{ij}^{soo}& = &- \frac{\alpha ^2}{2r_{ij}^3}\left( \left[
{\bf L}_{ij}+2{\bf L}_{ji}\right] \cdot {\bf s}_i\right)
\left( 1+P_{ij}\right) \nonumber \\
& = & -\frac{\alpha ^2}{2r_{ij}^3}\left( {\bf L}_{ij}\cdot \left[ {\bf s}_i+2{\bf s}_j\right] \right) \left(
1+P_{ij}\right) .
\end{eqnarray}
This, according to the definition (24-12) of~Slater \cite{S}
\begin{equation}
\label{eq:d}H^{soo}=\displaystyle
\sum_{i>j}H_{ij}^{soo},
\end{equation}
immediately yields the spin-other-orbit part of the Hamiltonian:

\begin{eqnarray}
\label{eq:aa}
H^{soo}& =& - \frac{\alpha ^2}2
\displaystyle \sum_{i>j}\frac 1{r_{ij}^3}\left( {\bf L}_{ij}\cdot \left[
{\bf s}_i+2{\bf s}_j\right] \right) \left( 1+P_{ij}\right) \nonumber \\
& = & -\frac{\alpha ^2}2\displaystyle \sum_{i\neq j}\frac 1{r_{ij}^3}\left(
{\bf L}_{ij}\cdot \left[ {\bf s}_i+2{\bf s}_j\right] \right) .
\end{eqnarray}
The expression (\ref{eq:aa}) in the formalism of second quantization is

\begin{equation}
\label{eq:e}\widehat{H}^{soo}=-\frac{\alpha ^2}2\sum_{iji^{\prime }j^{\prime
}}\left( ij\left| \frac{{\bf L}_{12}}{r_{12}^3}\cdot \left[ {\bf s}_1+2{\bf s%
}_2\right] \right| i^{\prime }j^{\prime }\right) a_ia_ja_{j^{\prime
}}^{\dagger }a_{i^{\prime }}^{\dagger },
\end{equation}
where the summation extends over all the possible single-electron states $%
iji^{\prime }j^{\prime }$ ($i\equiv n_il_im_{l_i}sm_{s_i}$ etc., $i$ and $%
i^{\prime }$ belong to the coordinate space of electron $1$, $j$ and $%
j^{\prime }$ - of electron $2$) instead of the numbered electrons $i$ and $j$%
. We denote electron creation operators by $a_i$, and annihilation operators
by hermitean conjugates, $a_i^{\dagger }$. The additional factor $1/2$
before the sum, usually occurring while passing over to the second
quantization representation (see {\it e.g.} (64.15) and (64.16) of~Landau
and Lifshitz \cite{LL}), is not present here because in (\ref{eq:aa}) not
only $i>j$, but also $j>i$ terms appear. However, the two-particle operator
(the one between bra and ket functions) in (\ref{eq:e}) is no longer
symmetric with respect to the permutation of electron labels $1$ and $2$.
One might symmetrize it, at the expense of doubling the number of
terms, which is unnecessary here. We should note also that actually in the
matrix element $\left( ij\left| \frac{{\bf L}_{12}}{r_{12}^3}\cdot \left[
{\bf s}_1+2{\bf s}_2\right] \right| i^{\prime }j^{\prime }\right) ${\it \ no
summation over spin indices is performed}, as such a summation should
include different spin indices in $a_ia_ja_{j^{\prime }}^{\dagger
}a_{i^{\prime }}^{\dagger }$, as well. Here some confusion may arise if we
accept the definitions presented in textbooks (even in~Landau and Lifshitz
\cite{LL}) too literally. However, we circumvent this point by using the
irreducible tensorial form of operator and the submatrix elements in radial
and spin-angular spaces, as we will show in the following section.

The operator between bra and ket functions on the right-hand side of (\ref
{eq:e}) is transformed to the irreducible tensorial form (in the spaces of
angular and spin momenta). We use the identity

\begin{equation}
\label{eq:f}\frac{{\bf L}_{12}}{r_{12}^3}=-\frac i{r_{12}^3}\left[ {\bf r}%
_{12}\times {\bf \nabla }_1\right] =i\left[ {\bf \nabla }_1\frac
1{r_{12}}\times {\bf \nabla }_1\right] ,
\end{equation}
or in tensorial form,

\begin{equation}
\label{eq:ff}\frac{L_{12}^{\quad (1)}}{r_{12}^3}=-i\sqrt{2}\left[ \nabla
_1^{\ (1)}\frac 1{r_{12}}\times \nabla _1^{\ (1)}\right] ^{(1)},
\end{equation}
\begin{equation}
\label{eq:g}\frac 1{r_{12}}=\sum_k\frac{r_{<}^k}{r_{>}^{k+1}}\left( C_1^{\
(k)}\cdot C_2^{\ (k)}\right) ,
\end{equation}
and

\begin{equation}
\label{eq:ii}\nabla _1^{\ (1)}=C_1^{\ (1)}\frac \partial {\partial r_1}+i%
\sqrt{2}\left[ C_1^{\ (1)}\times L_1^{\ (1)}\right] ^{(1)}
\end{equation}
of Jucys and Savukynas \cite{JS}, together with the commutator,

\begin{equation}
\label{eq:i}\left[ L_{1\rho }^{\ (1)},C_{1q}^{\ (k)}\right] =i\sqrt{k(k+1)}%
\left[
\begin{array}{ccc}
k & 1 & k \\
q & \rho & q+\rho
\end{array}
\right] C_{1q+\rho }^{\quad (k)},
\end{equation}
and the identity

\begin{equation}
\label{eq:jj}\left[ C_1^{\ (k+1)}\times L_1^{\ (1)}\right] ^{(k)}=-\sqrt{%
\frac{k(2k-1)}{(k+1)(2k+3)}}\left[ C_1^{\ (k-1)}\times L_1^{\ (1)}\right]
^{(k)}
\end{equation}
of~Kaniauskas and Rudzikas \cite{KR}, and the standard angular momenta
recoupling techniques, to obtain

$$
\frac{{\bf L}_{12}}{r_{12}^3}\cdot \left[ {\bf s}_1+2{\bf s}_2\right]
=\left( \frac{L_{12}^{\quad (1)}}{r_{12}^3}\cdot \left[ s_1^{\ (1)}+2s_2^{\
(1)}\right] \right)
$$
$$
=-\frac 1{\sqrt{3}}\displaystyle {\sum_k}\left\{ \left( \left[ \left[ C_1^{\
(k)}\times L_1^{\ (1)}\right] ^{(k-1)}\times C_2^{\ (k)}\right] ^{(1)}\cdot
\left[ s_1^{\ (1)}+2s_2^{\ (1)}\right] \right) \right.
$$
$$
\times \left( 2k+1\right) \sqrt{\left( 2k-1\right) }~\frac{r_1^{k-2}}{%
r_2^{k+1}}~\epsilon (r_2-r_1)
$$
$$
+\sqrt{\left( 2k+1\right) }\left( \left[ \left[ C_1^{\ (k)}\times L_1^{\
(1)}\right] ^{(k)}\times C_2^{\ (k)}\right] ^{(1)}\cdot \left[ s_1^{\
(1)}+2s_2^{\ (1)}\right] \right)
$$
$$
\times \left\{ \left( k+1\right) ~\frac{r_1^{k-2}}{r_2^{k+1}}~\epsilon
(r_2-r_1)-k~\frac{r_2^k}{r_1^{k+3}}~\epsilon (r_1-r_2)\right\}
$$
$$
-\left( \left[ C_1^{\ (k)}\times C_2^{\ (k)}\right] ^{(1)}\cdot \left[
s_1^{\ (1)}+2s_2^{\ (1)}\right] \right)
$$
$$
\times i\sqrt{k\left( k+1\right) \left( 2k+1\right) }~\frac{r_{<}^{k-1}}{%
r_{>}^{k+2}}~r_2~\frac \partial {\partial r_1}
$$
$$
-\left( \left[ \left[ C_1^{\ (k)}\times L_1^{\ (1)}\right] ^{(k+1)}\times
C_2^{\ (k)}\right] ^{(1)}\cdot \left[ s_1^{\ (1)}+2s_2^{\ (1)}\right]
\right)
$$
\begin{eqnarray}
\label{eq:m-a}
& & \left.  \times
\left(2k+1\right) \sqrt{\left( 2k+3\right) }
~\frac{r_2^k}{r_1^{k+3}}~ \epsilon (r_1-r_2)\right\} .
\end{eqnarray}
Here the tensorial operator of the spherical function is related to the
spherical function of~Condon and Shortley \cite{CS} by
\begin{equation}
\label{eq:h}C_{nq}^{\ (k)}=i^k\sqrt{\frac{4\pi }{2k+1}}Y(kq\mid \vartheta
_n\varphi _n)
\end{equation}
and $\epsilon (x)$ is a Heaviside step-function,

\begin{equation}
\label{eq:h-a}\epsilon (x)=\left\{
\begin{array}{ll}
1 ; & \mbox{ for } x>0, \\ 0 ; & \mbox{ for } x\leq 0.
\end{array}
\right.
\end{equation}

The expression (\ref{eq:m-a}) coincides with the one given in~Glass and
Hibbert \cite{GH} (formula (37), term for electrons $ij\equiv 12$), except
for the $i$ factor at the $\partial /\partial r_1$ term, which is missing
there. That irreducible tensorial form of the spin-other-orbit interaction
operator presented by
Glass and Hibbert \cite{GH} is perhaps the simplest known in the
literature, because it contains only six terms of different tensorial
structure, with only a single summation over the tensor ranks $k$. Here we
imply that a tensorial structure indexed by $(k_1k_2k,\sigma _1\sigma
_2\sigma )$ has rank $k_1$ for electron $1$, rank $k_2$ for electron $2$,
and a resulting rank $k$ in the $l$ space, and corresponding ranks $\sigma
_1\sigma _2\sigma $ in the $s$ space. Then in terms of different structures
we have

\begin{eqnarray}
\label{eq:m-b}
H_{12}^{soo} & \equiv & - \frac{\alpha
^2}2\left( \frac{{\bf L}_{12}}{r_{12}^3}\cdot \left[ {\bf s}_1+2
{\bf s}_2\right] \right) = \nonumber \\
& = & \displaystyle {\sum_k}\left\{
H_{soo}^{\left( k-1k1,101\right) } +H_{soo}^{\left( k-1k1,011\right)
}+H_{soo}^{\left( kk1,101\right) } \right.  \nonumber \\
& + & \left.
H_{soo}^{\left( kk1,011\right) }+H_{soo}^{\left( k+1k1,101\right)
}+H_{soo}^{\left( k+1k1,011\right) }\right\} ,
\end{eqnarray}
with

\begin{eqnarray}
\label{eq:m-c}
H_{soo}^{\left( k-1k1,101\right) }& = &
\frac{\alpha ^2}{2\sqrt{3}} \left( \left[ \left[ C_1^{\ (k)}\times
L_1^{\ (1)}\right] ^{(k-1)}\times C_2^{\ (k)}\right] ^{(1)}\cdot s_1^{\
(1)}\right) \nonumber \\
& \times & \left( 2k+1\right)
\sqrt{2k-1}~\frac{r_1^{k-2}}{r_2^{k+1}}~\epsilon (r_2-r_1),
\end{eqnarray}

\begin{eqnarray}
\label{eq:m-d}
H_{soo}^{\left( k-1k1,011\right) }& = &
\frac{\alpha ^2}{\sqrt{3}} \left( \left[ \left[ C_1^{\ (k)}\times L_1^{\
(1)}\right] ^{(k-1)}\times C_2^{\ (k)}\right] ^{(1)}\cdot s_2^{\
(1)}\right) \nonumber \\
& \times & \left( 2k+1\right)
\sqrt{2k-1}~\frac{r_1^{k-2}}{r_2^{k+1}}~\epsilon (r_2-r_1),
\end{eqnarray}

\begin{eqnarray}
\label{eq:m-e}
H_{soo}^{\left(
kk1,101\right) } & = & \frac{\alpha ^2}{2\sqrt{3}}\left\{ \left(
\sqrt{2k+1}
\left[
\left[ C_1^{\ (k)}\times L_1^{\ (1)}\right] ^{(k)}\times C_2^{\
(k)}\right] ^{(1)}\cdot s_1^{\ (1)}\right) \right.  \nonumber \\
& \times &
\left\{ \left( k+1\right) ~\frac{r_1^{k-2}}{r_2^{k+1}}~ \epsilon
(r_2-r_1)-k ~\frac{r_2^k}{r_1^{k+3}}~ \epsilon (r_1-r_2)\right\}
\nonumber \\
 & - &  i \sqrt{k\left( k+1\right) \left( 2k+1\right)}
 \left( \left[ C_1^{\ (k)}\times C_2^{\ (k)}
\right] ^{(1)}\cdot s_1^{\ (1)}\right)
\nonumber \\
& \times &
\left. ~\frac{r_{<}^{k-1}}{ r_{>}^{k+2}}~r_2~\frac \partial {\partial
r_1}\right\},
\end{eqnarray}

\begin{eqnarray}
\label{eq:m-f}
H_{soo}^{\left(
kk1,011\right) } & = & \frac{\alpha ^2}{\sqrt{3}}\left\{ \sqrt{2k+1}
\left( \left[ \left[
C_1^{\ (k)}\times L_1^{\ (1)}\right] ^{(k)}\times C_2^{\ (k)}\right]
^{(1)}\cdot s_2^{\ (1)}\right) \right.  \nonumber \\
& \times & \left\{
\left( k+1\right)~ \frac{r_1^{k-2}}{r_2^{k+1}}~ \epsilon(r_2-r_1)
-k~ \frac{r_2^k}{r_1^{k+3}}~ \epsilon (r_1-r_2)\right\} \nonumber \\
& - & i
\sqrt{k\left( k+1\right) \left( 2k+1\right)}
 \left( \left[ C_1^{\ (k)}\times C_2^{\ (k)}\right] ^{(1)}\cdot s_2^{\
(1)}\right) \nonumber \\
& \times &
\left. \frac{r_{<}^{k-1}}{ r_{>}^{k+2}}~r_2~\frac \partial {\partial
r_1}\right\},
\end{eqnarray}

\begin{eqnarray}
\label{eq:m-g}
H_{soo}^{\left( k+1k1,101\right) }& = &
- \frac{\alpha ^2}{2\sqrt{3}}\left( \left[ \left[ C_1^{\ (k)}\times
L_1^{\ (1)}\right] ^{(k+1)}\times C_2^{\ (k)}\right] ^{(1)}\cdot s_1^{\
(1)}\right) \nonumber \\
& \times & \left( 2k+1\right) \sqrt{2k+3}
~\frac{r_2^k}{r_1^{k+3}}~ \epsilon (r_1-r_2),
\end{eqnarray}

\begin{eqnarray}
\label{eq:m-h}
H_{soo}^{\left( k+1k1,011\right) } & = & -
\frac{\alpha ^2}{\sqrt{3}}\left( \left[ \left[ C_1^{\ (k)}\times L_1^{\
(1)}\right] ^{(k+1)}\times C_2^{\ (k)}\right] ^{(1)}\cdot s_2^{\ (1)}\right)
\nonumber \\
& \times & \left( 2k+1\right) \sqrt{2k+3} ~\frac{r_2^k}{r_1^{k+3}}~ \epsilon
(r_1-r_2),
\end{eqnarray}
where the tensor ranks $k$ for (\ref{eq:m-c})-(\ref{eq:m-f}) satisfy the
condition $k\geq 1$ and for (\ref{eq:m-g})-(\ref{eq:m-h}) $k\geq 0$.

Now, since we have from (\ref{eq:e}) that

\begin{equation}
\label{eq:k}\widehat{H}^{soo}=\displaystyle \sum_{iji^{\prime }j^{\prime
}}\left( ij\left| H_{12}^{soo}\right| i^{\prime }j^{\prime }\right)
a_ia_ja_{j^{\prime }}^{\dagger }a_{i^{\prime }}^{\dagger },
\end{equation}
we readily obtain the expressions for particular terms $\widehat{H}%
_{soo}^{(k_1k_2k,\sigma _1\sigma _2\sigma )}$ of $\widehat{H}^{soo}$ in a
coupled tensorial form from (7) or (8) of P1~\cite{GR} by
taking $\widehat{H}_{soo}^{(k_1k_2k,\sigma _1\sigma _2\sigma )}$ for $G$ and
$H_{soo}^{(k_1k_2k,\sigma _1\sigma _2\sigma )}$ for $\frac 12g$ there. Those
operators in the formalism of second quantization are further transformed to
arrive at the form schematically outlined in (5)-(8) of
P2~\cite{GRF} (with $\alpha$, $\beta$, $\gamma$, $\delta$ being strictly
different) as

\begin{equation}
\label{eq:m-i}
\begin{array}[b]{c}
\widehat{G}\sim
\displaystyle {\sum_{\alpha}}
\displaystyle {\sum_{\kappa _{12},\sigma _{12},\kappa
_{12}^{\prime },\sigma _{12}^{\prime }}}\Theta \left( \Xi \right) \left\{
A_{p,-p}^{\left( kk\right) }\left( n_\alpha \lambda _\alpha ,\Xi \right)
\delta \left( u,1\right) \right. \\
+
\displaystyle {\sum_{\beta}}
\left[ B^{\left( \kappa _{12}\sigma
_{12}\right) }\left( n_\alpha \lambda _\alpha ,\Xi \right) \times C^{\left(
\kappa _{12}^{\prime }\sigma _{12}^{\prime }\right) }\left( n_\beta \lambda
_\beta ,\Xi \right) \right] _{p,-p}^{\left( kk\right) }\delta \left(
u,2\right) \\
+
\displaystyle {\sum_{\beta \gamma}}
\left[ \left[ D^{\left( l_\alpha s\right) }\times D^{\left( l_\beta
s\right) }\right] ^{\left( \kappa _{12}\sigma _{12}\right) }\times E^{\left(
\kappa _{12}^{\prime }\sigma _{12}^{\prime }\right) }\left( n_\gamma \lambda
_\gamma ,\Xi \right) \right] _{p,-p}^{\left( kk\right) }\delta \left(
u,3\right) \\
\left. +
\displaystyle {\sum_{\beta \gamma \delta}}
\left[ \left[ D^{\left( l_\alpha s\right) }\times D^{\left( l_\beta
s\right) }\right] ^{\left( \kappa _{12}\sigma _{12}\right) }\times \left[
D^{\left( l_\gamma s\right) }\times D^{\left( l_\delta s\right) }\right]
^{\left( \kappa _{12}^{\prime }\sigma _{12}^{\prime }\right) }\right]
_{p,-p}^{\left( kk\right) }\delta \left( u,4\right) \right\} .
\end{array}
\end{equation}
Here $A^{\left( kk\right) }\left( n\lambda ,\Xi \right) ,...,E^{\left(
kk^{\prime }\right) }\left( n\lambda ,\Xi \right) $ denote tensorial
products of those creation/annihilation operators that act upon a particular
electron shell (see P2~\cite{GRF}), $\lambda
\equiv ls$, and $u$ is the overall number of shells acted upon by a given
tensorial product of creation/annihilation operators. Parameter $\Xi $
implies the whole array of parameters (and sometimes an internal summation
over some of these is implied, as well) that connect the amplitudes $\Theta $
of tensorial products of creation/annihilation operators in the expression (%
\ref{eq:m-i}) to these tensorial products (see P2~\cite{GRF}).
These amplitudes $\Theta \left( \Xi \right) $ are all
proportional to the submatrix element of a two-particle operator $g$,

\begin{equation}
\label{eq:m-ia}\Theta \left( \Xi \right) \sim \left( n_i\lambda _in_j\lambda
_j\left\| g\right\| n_{i^{\prime }}\lambda _{i^{\prime }}n_{j^{\prime
}}\lambda _{j^{\prime }}\right) .
\end{equation}

In the following section we present the explicit expressions of submatrix
elements for particular terms of $H_{12}^{soo}$ defined by (\ref{eq:m-b}).

\section{Submatrix Elements for the Spin-Other-Orbit Operator Amplitudes}

There are six terms having different tensorial structure, summed over $k$ in
$H_{12}^{soo}$ expansion (\ref{eq:m-b}). Their submatrix elements are all
contained in the following three expressions, provided the appropriate $%
\sigma _1$ and $\sigma _2$ are chosen:

\begin{equation}
\label{eq:b-b}
\begin{array}[b]{c}
\left( n_i\lambda _in_j\lambda _j\left\| H_{soo}^{\left( k-1k1,\sigma
_1\sigma _21\right) }\right\| n_{i^{\prime }}\lambda _{i^{\prime
}}n_{j^{\prime }}\lambda _{j^{\prime }}\right) =2\cdot 2^{\sigma _2}\left\{
\left( 2k-1\right) \left( 2k+1\right) \right. \\
\times \left. \left( l_i+l_{i^{\prime }}-k+1\right) \left(
k-l_i+l_{i^{\prime }}\right) \left( k+l_i-l_{i^{\prime }}\right) \left(
k+l_i+l_{i^{\prime }}+1\right) \right\} ^{1/2} \\
\times \left( k\right) ^{-1/2}\left( l_i\left\| C^{\left( k\right) }\right\|
l_{i^{\prime }}\right) \left( l_j\left\| C^{\left( k\right) }\right\|
l_{j^{\prime }}\right) N^{k-2}\left( n_jl_jn_il_i,n_{j^{\prime
}}l_{j^{\prime }}n_{i^{\prime }}l_{i^{\prime }}\right) ;
\end{array}
\end{equation}

\begin{equation}
\label{eq:b-c}
\begin{array}[b]{c}
\left( n_i\lambda _in_j\lambda _j\left\| H_{soo}^{\left( kk1,\sigma _1\sigma
_21\right) }\right\| n_{i^{\prime }}\lambda _{i^{\prime }}n_{j^{\prime
}}\lambda _{j^{\prime }}\right) =-2\cdot 2^{\sigma _2}\left( 2k+1\right)
^{1/2}\left( l_i\left\| C^{\left( k\right) }\right\| l_{i^{\prime }}\right)
\\
\times \left( l_j\left\| C^{\left( k\right) }\right\| l_{j^{\prime }}\right)
\left\{ \left( k\left( k+1\right) \right) ^{-1/2}\left( l_i\left(
l_i+1\right) -k\left( k+1\right) -l_{i^{\prime }}\left( l_{i^{\prime
}}+1\right) \right) \right. \\
\times \left\{ \left( k+1\right) N^{k-2}\left( n_jl_jn_il_i,n_{j^{\prime
}}l_{j^{\prime }}n_{i^{\prime }}l_{i^{\prime }}\right) -kN^k\left(
n_il_in_jl_j,n_{i^{\prime }}l_{i^{\prime }}n_{j^{\prime }}l_{j^{\prime
}}\right) \right\} \\
\left. -2\left( k\left( k+1\right) \right) ^{1/2}V^{k-1}\left(
n_il_in_jl_j,n_{i^{\prime }}l_{i^{\prime }}n_{j^{\prime }}l_{j^{\prime
}}\right) \right\} ;
\end{array}
\end{equation}

\begin{equation}
\label{eq:b-d}
\begin{array}[b]{c}
\left( n_i\lambda _in_j\lambda _j\left\| H_{soo}^{\left( k+1k1,\sigma
_1\sigma _21\right) }\right\| n_{i^{\prime }}\lambda _{i^{\prime
}}n_{j^{\prime }}\lambda _{j^{\prime }}\right) =2\cdot 2^{\sigma _2}\left\{
\left( 2k+1\right) \left( 2k+3\right) \right. \\
\times \left. \left( l_i+l_{i^{\prime }}-k\right) \left( k-l_i+l_{i^{\prime
}}+1\right) \left( k+l_i-l_{i^{\prime }}+1\right) \left( k+l_i+l_{i^{\prime
}}+2\right) \right\} ^{1/2} \\
\times \left( k+1\right) ^{-1/2}\left( l_i\left\| C^{\left( k\right)
}\right\| l_{i^{\prime }}\right) \left( l_j\left\| C^{\left( k\right)
}\right\| l_{j^{\prime }}\right) N^k\left( n_il_in_jl_j,n_{i^{\prime
}}l_{i^{\prime }}n_{j^{\prime }}l_{j^{\prime }}\right) .
\end{array}
\end{equation}
The radial integrals of two types occurring in (\ref{eq:b-b})-(\ref{eq:b-d})
are (see, Glass and Hibbert \cite{GH}):

$$
N^k\left( n_il_in_jl_j,n_{i^{\prime }}l_{i^{\prime }}n_{j^{\prime
}}l_{j^{\prime }}\right)%
$$
\begin{eqnarray}
\label{eq:m-j}
=\frac{\alpha ^2}4\int_0^\infty \int_0^\infty P_i\left( r_1\right) P_j\left(
r_2\right) \frac{r_2^k}{r_1^{k+3}}\epsilon (r_1-r_2)P_{i^{\prime }}\left(
r_1\right) P_{j^{\prime }}\left( r_2\right) dr_1dr_{2,}
\end{eqnarray}
and

$$
V^k\left( n_il_in_jl_j,n_{i^{\prime }}l_{i^{\prime }}n_{j^{\prime
}}l_{j^{\prime }}\right)
$$
\begin{eqnarray}
\label{eq:m-k}
=\frac{\alpha ^2}4\int_0^\infty \int_0^\infty P_i\left( r_1\right) P_j\left(
r_2\right) \frac{r_{<}^{k-1}}{r_{>}^{k+2}}r_2\frac \partial {\partial
r_1}P_{i^{\prime }}\left( r_1\right) P_{j^{\prime }}\left( r_2\right)
dr_1dr_2.
\end{eqnarray}
The integrals $N^k\left( n_il_in_jl_j,n_{i^{\prime }}l_{i^{\prime
}}n_{j^{\prime }}l_{j^{\prime }}\right) $ have the following symmetry
properties:

\begin{equation}
\label{eq:m-ka}
\begin{array}[b]{c}
N^k\left( n_il_in_jl_j,n_{i^{\prime }}l_{i^{\prime }}n_{j^{\prime
}}l_{j^{\prime }}\right) =N^k\left( n_{i^{\prime }}l_{i^{\prime
}}n_{j^{\prime }}l_{j^{\prime }},n_il_in_jl_j\right) \\
=N^k\left( n_{i^{\prime }}l_{i^{\prime }}n_jl_j,n_il_in_{j^{\prime
}}l_{j^{\prime }}\right) =N^k\left( n_il_in_{j^{\prime }}l_{j^{\prime
}},n_{i^{\prime }}l_{i^{\prime }}n_jl_j\right) .
\end{array}
\end{equation}
As was shown in the monograph of Jucys and Savukynas \cite{JS}, and later
in~the paper of Godefroid \cite{Gf}, the integrals $N^k\left(
n_il_in_jl_j,n_{i^{\prime }}l_{i^{\prime }}n_{j^{\prime }}l_{j^{\prime
}}\right) $ and $V^k\left( n_il_in_jl_j,n_{i^{\prime }}l_{i^{\prime
}}n_{j^{\prime }}l_{j^{\prime }}\right) $ are related by
\begin{equation}
\label{eq:m-l}
\begin{array}[b]{c}
V^{k-1}\left( n_il_in_jl_j,n_{i^{\prime }}l_{i^{\prime }}n_{j^{\prime
}}l_{j^{\prime }}\right) +V^{k-1}\left( n_{i^{\prime }}l_{i^{\prime
}}n_{j^{\prime }}l_{j^{\prime }},n_il_in_jl_j\right) \\
=kN^k\left( n_il_in_jl_j,n_{i^{\prime }}l_{i^{\prime }}n_{j^{\prime
}}l_{j^{\prime }}\right) -\left( k+1\right) N^{k-2}\left(
n_jl_jn_il_i,n_{j^{\prime }}l_{j^{\prime }}n_{i^{\prime }}l_{i^{\prime
}}\right) .
\end{array}
\end{equation}

The use of the approach presented in P2~\cite{GRF}
presumes that both the tensorial structure of th thee operator
under consideration and the submatrix elements $\left( n_i\lambda
_in_j\lambda _j\left\| g\right\| n_{i^{\prime }}\lambda _{i^{\prime
}}n_{j^{\prime }}\lambda _{j^{\prime }}\right) $ are known. The formulae (%
\ref{eq:b-b}), (\ref{eq:b-c}) and (\ref{eq:b-d}) are the expressions we
need, with the fixed tensorial structures of $H_{soo}^{\left( \kappa
_1\kappa _2\kappa ,\sigma _1\sigma _2\sigma \right) }$, corresponding to $%
\frac 12g^{\left( \kappa _1\kappa _2\kappa ,\sigma _1\sigma _2\sigma \right)
}$ of a general operator of P2~\cite{GRF} (we
could use $H_{12}^{soo}+H_{21}^{soo}$, which would correspond just to $g$,
but that is unnecessary, as stated earlier). We may readily obtain the value
of a matrix element of this operator for any number of open shells in bra
and ket functions, by choosing every tensorial structure from (\ref{eq:m-b}%
), using their submatrix elements and corresponding tensorial ranks in an
expression of the type (\ref{eq:m-i}), defining bra and ket functions, and
performing spin-angular integrations according to P2~\cite{GRF}.

\section{Some Simplifications for Submatrix Elements}

In this section we will discuss some special cases of distributions
$iji^{\prime }j^{\prime }$ for the spin-other-orbit interaction operator.
The labels $iji^{\prime }j^{\prime }$ in the expressions starting from
(\ref{eq:e}), and then
(\ref{eq:k}) and further, do not necessarily label the different
single-electron states (although some combinations cancel in second quantized
expressions (\ref{eq:e}) and (\ref{eq:k}): only $i \neq j$ and $ i^{\prime} \neq j^{\prime }$
terms remain). Now we will use strictly different indices $\alpha$ and
$\beta$, introduced in P2~\cite{GRF} (see Table 1 there), to distinguish between
separate cases of the coinciding principle and angular momentum quantum numbers
$n$ and $\lambda$ in the arrays $iji^{\prime }j^{\prime }$,
$i \equiv n_i l_i s_i m_{l_i} m_{s_i}$.
In these cases of coincidence some of the submatrix elements vanish,
and therefore can be omitted
in spin-angular integrations, thus simplifying the calculations.

\subsection{Distribution $iji^{\prime }j^{\prime }=\alpha \alpha \alpha
\alpha $}

For the distribution $iji^{\prime }j^{\prime }=\alpha \alpha \alpha \alpha $%
, on the basis of the relation for radial integrals (\ref{eq:m-l}), we
easily see that those integrals compensate each other in tensorial
structures $(kk1,101)$ and $(kk1,011)$:

\begin{equation}
\label{eq:m-la}
\begin{array}[b]{c}
\left( n_\alpha \lambda _\alpha n_\alpha \lambda _\alpha \left\|
H_{soo}^{\left( kk1,\sigma _1\sigma _21\right) }\right\| n_\alpha \lambda
_\alpha n_\alpha \lambda _\alpha \right) =-2\cdot 2^{\sigma _2}\left(
2k+1\right) ^{1/2} \\
\times \left( l_\alpha \left\| C^{\left( k\right) }\right\| l_\alpha \right)
^2\left\{ \left( k\left( k+1\right) \right) ^{-1/2}\left( l_\alpha \left(
l_\alpha +1\right) -k\left( k+1\right) -l_\alpha \left( l_\alpha +1\right)
\right) \right. \\
\times \left\{ \left( k+1\right) N^{k-2}\left( n_\alpha l_\alpha n_\alpha
l_\alpha ,n_\alpha l_\alpha n_\alpha l_\alpha \right) -kN^k\left( n_\alpha
l_\alpha n_\alpha l_\alpha ,n_\alpha l_\alpha n_\alpha l_\alpha \right)
\right\} \\
\left. -2\left( k\left( k+1\right) \right) ^{1/2}V^{k-1}\left( n_\alpha
l_\alpha n_\alpha l_\alpha ,n_\alpha l_\alpha n_\alpha l_\alpha \right)
\right\} =0.
\end{array}
\end{equation}
Then from expressions (47), (48) and (49) of P2~\cite{GRF},
and using expression (\ref{eq:b-c}) for $H_{12}^{soo}$, we
obtain the final tensorial form of spin-other-orbit interaction operator
acting within a particular shell of electrons $\alpha $:

\begin{equation}
\label{eq:b-ca}
\begin{array}[b]{c}
\widehat{H}_{12}^{soo}(\alpha \alpha \alpha \alpha )=\displaystyle {\sum_k}%
\displaystyle {\sum_p}(-1)^{1-p}\left\{ \left( n_\alpha \lambda _\alpha
n_\alpha \lambda _\alpha \left\| H_{soo}^{\left( k-1k1,101\right) }\right\|
n_\alpha \lambda _\alpha n_\alpha \lambda _\alpha \right) \right. \\ \times
[k-1,k,1]^{-1/2}\left[ \left[ a^{\left( l_\alpha s\right) }\times \tilde
a^{\left( l_\alpha s\right) }\right] ^{\left( k-11\right) }\times \left[
a^{\left( l_\alpha s\right) }\times \tilde a^{\left( l_\alpha s\right)
}\right] ^{\left( k0\right) }\right] _{p,-p}^{\left( 11\right) } \\
+\left( n_\alpha \lambda _\alpha n_\alpha \lambda _\alpha \left\|
H_{soo}^{\left( k-1k1,011\right) }\right\| n_\alpha \lambda _\alpha n_\alpha
\lambda _\alpha \right) \\
\times [k-1,k,1]^{-1/2}\left[ \left[ a^{\left( l_\alpha s\right) }\times
\tilde a^{\left( l_\alpha s\right) }\right] ^{\left( k-10\right) }\times
\left[ a^{\left( l_\alpha s\right) }\times \tilde a^{\left( l_\alpha
s\right) }\right] ^{\left( k1\right) }\right] _{p,-p}^{\left( 11\right) } \\
+\left( n_\alpha \lambda _\alpha n_\alpha \lambda _\alpha \left\|
H_{soo}^{\left( k+1k1,101\right) }\right\| n_\alpha \lambda _\alpha n_\alpha
\lambda _\alpha \right) \\
\times [k+1,k,1]^{-1/2}\left[ \left[ a^{\left( l_\alpha s\right) }\times
\tilde a^{\left( l_\alpha s\right) }\right] ^{\left( k+11\right) }\times
\left[ a^{\left( l_\alpha s\right) }\times \tilde a^{\left( l_\alpha
s\right) }\right] ^{\left( k0\right) }\right] _{p,-p}^{\left( 11\right) } \\
+\left( n_\alpha \lambda _\alpha n_\alpha \lambda _\alpha \left\|
H_{soo}^{\left( k+1k1,011\right) }\right\| n_\alpha \lambda _\alpha n_\alpha
\lambda _\alpha \right) \\
\times [k+1,k,1]^{-1/2}\left[ \left[ a^{\left( l_\alpha s\right) }\times
\tilde a^{\left( l_\alpha s\right) }\right] ^{\left( k+10\right) }\times
\left[ a^{\left( l_\alpha s\right) }\times \tilde a^{\left( l_\alpha
s\right) }\right] ^{\left( k1\right) }\right] _{p,-p}^{\left( 11\right) } \\
-\left\{ \left( k(2l_\alpha -k+1)(2l_\alpha +k+1)\right)
^{1/2}[k-1]^{-1/2}\right. \\
\times \left( n_\alpha \lambda _\alpha n_\alpha \lambda _\alpha \left\|
H_{soo}^{\left( k-1k1,101\right) }\right\| n_\alpha \lambda _\alpha n_\alpha
\lambda _\alpha \right) \\
-\left( (k+1)(2l_\alpha -k)(2l_\alpha +k+2)\right) ^{1/2}[k+1]^{-1/2} \\
\times \left. \left( n_\alpha \lambda _\alpha n_\alpha \lambda _\alpha
\left\| H_{soo}^{\left( k+1k1,101\right) }\right\| n_\alpha \lambda _\alpha
n_\alpha \lambda _\alpha \right) \right\} \\
\times \left. \frac 12\sqrt{\frac 32}(-1)^k\left( l_\alpha (l_\alpha
+1)[l_\alpha ,k]\right) ^{-1/2}\left[ a^{\left( l_\alpha s\right) }\times
\tilde a^{\left( l_\alpha s\right) }\right] _{p,-p}^{\left( 11\right)
}\right\} .
\end{array}
\end{equation}
We define the tensor $\tilde a^{\left( ls\right) }$ as related to the
electron annihilation operator $a_{-m_l,-m_s}^{\left( l\ s\right) \dagger }$
by Rudzikas \cite{R},

\begin{equation}
\label{eq:b-ca-a}\tilde a_{m_lm_s}^{\left( l\ s\right) }=\left( -1\right)
^{l+s-m_l-m_s}a_{-m_l,-m_s}^{\left( l\ s\right) \dagger }
\end{equation}
and use a shorthand notation $(2k+1)\cdot ...\equiv [k,...]$ .

We also have from (\ref{eq:b-b}) and (\ref{eq:b-d}):

\begin{equation}
\label{eq:b-caa}
\begin{array}[b]{c}
\left( n_\alpha \lambda _\alpha n_\alpha \lambda _\alpha \left\|
H_{soo}^{\left( k-1k1,\sigma _1\sigma _21\right) }\right\| n_\alpha \lambda
_\alpha n_\alpha \lambda _\alpha \right) =2\cdot 2^{\sigma _2} \\
\times \left( \left( 2k-1\right) k\left( 2k+1\right) \left( 2l_\alpha
-k+1\right) \left( 2l_\alpha +k+1\right) \right) ^{1/2} \\
\times \left( l_\alpha \left\| C^{\left( k\right) }\right\| l_\alpha \right)
^2N^{k-2}\left( n_\alpha l_\alpha n_\alpha l_\alpha ,n_\alpha l_\alpha
n_\alpha l_\alpha \right)
\end{array}
\end{equation}

\begin{equation}
\label{eq:b-cab}
\begin{array}[b]{c}
\left( n_\alpha \lambda _\alpha n_\alpha \lambda _\alpha \left\|
H_{soo}^{\left( k+1k1,\sigma _1\sigma _21\right) }\right\| n_\alpha \lambda
_\alpha n_\alpha \lambda _\alpha \right) =2\cdot 2^{\sigma _2} \\
\times \left( \left( 2k+1\right) \left( k+1\right) \left( 2k+3\right) \left(
2l_\alpha -k\right) \left( 2l_\alpha +k+2\right) \right) ^{1/2} \\
\times \left( l_\alpha \left\| C^{\left( k\right) }\right\| l_\alpha \right)
^2N^k\left( n_\alpha l_\alpha n_\alpha l_\alpha ,n_\alpha l_\alpha n_\alpha
l_\alpha \right) .
\end{array}
\end{equation}
An expression equivalent to (\ref{eq:b-ca}) (with (\ref{eq:b-caa}) and (\ref
{eq:b-cab})) was already presented in the monograph Jucys and Savukynas \cite
{JS}, formulae (13.23) and (13.24), where a matrix element of
spin-other-orbit interaction within a single shell of equivalent electrons
is defined. The differences are that they use the coordinate representation,
and the Marvin notation of radial integrals (see~Marvin \cite{M}), where

\begin{equation}
\label{eq:b-cb}M_k\left( n_il_i,n_jl_j\right) =N^k\left(
n_il_in_jl_j,n_il_in_jl_j\right) .
\end{equation}

Thus, there are four terms $H_{soo}^{\left( k-1k1,101\right) }$, $%
H_{soo}^{\left( k-1k1,011\right) }$, $H_{soo}^{\left( k+1k1,101\right) }$
and $H_{soo}^{\left( k+1k1,011\right) }$ having different tensorial
structure for this distribution instead of six (see expression (\ref{eq:m-b}%
)). All of them are general in the sense that they may be applied to obtain
matrix elements of spin-other-orbit interaction operator for distribution $%
\alpha \alpha \alpha \alpha $ between functions with any number of open
electronic shells (see P2~\cite{GRF}).

\subsection{Distributions $iji^{\prime }j^{\prime }=\alpha \beta \alpha
\beta $ and $\beta \alpha \beta \alpha $}

For the distributions $iji^{\prime }j^{\prime }=\alpha \beta \alpha \beta $
and $\beta \alpha \beta \alpha $ we also have that the submatrix elements $%
\left( n_\alpha \lambda _\alpha n_\beta \lambda _\beta \left\|
H_{soo}^{\left( kk1,\sigma _1\sigma _21\right) }\right\| n_\alpha \lambda
_\alpha n_\beta \lambda _\beta \right) $ and

$\left( n_\beta \lambda _\beta n_\alpha \lambda _\alpha \left\|
H_{soo}^{\left( kk1,\sigma _1\sigma _21\right) }\right\| n_\beta \lambda
_\beta n_\alpha \lambda _\alpha \right) $ vanish, on the basis of the same
relation (\ref{eq:m-l}). Then from expressions (50) and (51) of P2~\cite{GRF},
we obtain the final tensorial form of the
spin-other-orbit interaction operator for the distribution $\alpha \beta
\alpha \beta $:

\begin{equation}
\label{eq:b-dd}
\begin{array}[b]{c}
\widehat{H}_{12}^{soo}(\alpha \beta \alpha \beta )=\displaystyle {\sum_k}%
\displaystyle {\sum_p}(-1)^{1-p} \left\{ \left( n_\alpha \lambda _\alpha
n_\beta \lambda _\beta \left\| H_{soo}^{\left( k-1k1,101\right) }\right\|
n_\alpha \lambda _\alpha n_\beta \lambda _\beta \right) \right. \\ \times
[k-1,k,1]^{-1/2}\left[ \left[ a^{\left( l_\alpha s\right) }\times \tilde
a^{\left( l_\alpha s\right) }\right] ^{\left( k-11\right) }\times \left[
a^{\left( l_\beta s\right) }\times \tilde a^{\left( l_\beta s\right)
}\right] ^{\left( k0\right) }\right] _{p,-p}^{\left( 11\right) } \\
+\left( n_\alpha \lambda _\alpha n_\beta \lambda _\beta \left\|
H_{soo}^{\left( k-1k1,011\right) }\right\| n_\alpha \lambda _\alpha n_\beta
\lambda _\beta \right) \times \\
\times [k-1,k,1]^{-1/2}\left[ \left[ a^{\left( l_\alpha s\right) }\times
\tilde a^{\left( l_\alpha s\right) }\right] ^{\left( k-10\right) }\times
\left[ a^{\left( l_\beta s\right) }\times \tilde a^{\left( l_\beta s\right)
}\right] ^{\left( k1\right) }\right] _{p,-p}^{\left( 11\right) } \\
+\left( n_\alpha \lambda _\alpha n_\beta \lambda _\beta \left\|
H_{soo}^{\left( k+1k1,101\right) }\right\| n_\alpha \lambda _\alpha n_\beta
\lambda _\beta \right) \times \\
\times [k+1,k,1]^{-1/2}\left[ \left[ a^{\left( l_\alpha s\right) }\times
\tilde a^{\left( l_\alpha s\right) }\right] ^{\left( k+11\right) }\times
\left[ a^{\left( l_\beta s\right) }\times \tilde a^{\left( l_\beta s\right)
}\right] ^{\left( k0\right) }\right] _{p,-p}^{\left( 11\right) } \\
+\left( n_\alpha \lambda _\alpha n_\beta \lambda _\beta \left\|
H_{soo}^{\left( k+1k1,011\right) }\right\| n_\alpha \lambda _\alpha n_\beta
\lambda _\beta \right) \times \\
\times \left. [k+1,k,1]^{-1/2}\left[ \left[ a^{\left( l_\alpha s\right) }
\times
\tilde a^{\left( l_\alpha s\right) }\right] ^{\left( k+10\right) }\times
\left[ a^{\left( l_\beta s\right) }\times \tilde a^{\left( l_\beta s\right)
}\right] ^{\left( k1\right) }\right] _{p,-p}^{\left( 11\right) } \right\}
\end{array}
\end{equation}
and for the distribution $\beta \alpha \beta \alpha $:

\begin{equation}
\label{eq:b-e}
\begin{array}[b]{c}
\begin{array}[b]{c}
\widehat{H}_{12}^{soo}(\beta \alpha \beta \alpha )=\displaystyle {\sum_k}%
\displaystyle {\sum_p}(-1)^{1-p} \left\{ \left( n_\beta \lambda _\beta
n_\alpha \lambda _\alpha \left\| H_{soo}^{\left( k-1k1,101\right) }\right\|
n_\beta \lambda _\beta n_\alpha \lambda _\alpha \right) \right. \\ \times
[k-1,k,1]^{-1/2}\left[ \left[ a^{\left( l_\alpha s\right) }\times \tilde
a^{\left( l_\alpha s\right) }\right] ^{\left( k0\right) }\times \left[
a^{\left( l_\beta s\right) }\times \tilde a^{\left( l_\beta s\right)
}\right] ^{\left( k-11\right) }\right] _{p,-p}^{\left( 11\right) } \\
+\left( n_\beta \lambda _\beta n_\alpha \lambda _\alpha \left\|
H_{soo}^{\left( k-1k1,011\right) }\right\| n_\beta \lambda _\beta n_\alpha
\lambda _\alpha \right) \\
\times [k-1,k,1]^{-1/2}\left[ \left[ a^{\left( l_\alpha s\right) }\times
\tilde a^{\left( l_\alpha s\right) }\right] ^{\left( k1\right) }\times
\left[ a^{\left( l_\beta s\right) }\times \tilde a^{\left( l_\beta s\right)
}\right] ^{\left( k-10\right) }\right] _{p,-p}^{\left( 11\right) } \\
+\left( n_\beta \lambda _\beta n_\alpha \lambda _\alpha \left\|
H_{soo}^{\left( k+1k1,101\right) }\right\| n_\beta \lambda _\beta n_\alpha
\lambda _\alpha \right) \\
\times [k+1,k,1]^{-1/2}\left[ \left[ a^{\left( l_\alpha s\right) }\times
\tilde a^{\left( l_\alpha s\right) }\right] ^{\left( k0\right) }\times
\left[ a^{\left( l_\beta s\right) }\times \tilde a^{\left( l_\beta s\right)
}\right] ^{\left( k+11\right) }\right] _{p,-p}^{\left( 11\right) } \\
+\left( n_\beta \lambda _\beta n_\alpha \lambda _\alpha \left\|
H_{soo}^{\left( k+1k1,011\right) }\right\| n_\beta \lambda _\beta n_\alpha
\lambda _\alpha \right) \\
\times \left. [k+1,k,1]^{-1/2}\left[ \left[ a^{\left( l_\alpha s\right) }
\times
\tilde a^{\left( l_\alpha s\right) }\right] ^{\left( k1\right) }\times
\left[ a^{\left( l_\beta s\right) }\times \tilde a^{\left( l_\beta s\right)
}\right] ^{\left( k+10\right) }\right] _{p,-p}^{\left( 11\right) } \right\}.
\end{array}
\end{array}
\end{equation}

The expression (\ref{eq:b-e}) can be obtained from (\ref{eq:b-dd})
by interchange $\alpha \rightleftharpoons \beta$  and anticommutation of the
second quantization operators. We present it here because according
to the approach of P2~\cite {GRF} the condition $\alpha < \beta$ is
imposed upon $\alpha$, $\beta$, so the distributions $\alpha \beta \alpha
\beta$ and $\beta \alpha \beta \alpha$ are different.

We obtain the submatrix elements appearing in (\ref{eq:b-dd}) and (\ref
{eq:b-e}) from (\ref{eq:b-b}) and (\ref{eq:b-d}). In these two cases the
tensorial form of the spin-other-orbit interaction operator also contains
the radial integrals of only one type, as in (\ref{eq:b-ca}), i.e. $%
N^k\left( n_il_in_jl_j,n_{i^{\prime }}l_{i^{\prime }}n_{j^{\prime
}}l_{j^{\prime }}\right) $. These tensorial forms (\ref{eq:b-dd}) and (\ref
{eq:b-e}) are general in the sense that they may be applied to obtain matrix
elements for given distributions between functions with any number of open
electronic shells, as stated already in P2~\cite {GRF}.
Then the case of just two open electronic shells would be a special
one, and it was treated by \cite{JS}. Those authors had obtained expressions
for matrix elements of direct interaction terms diagonal with respect to
configuration, containing one type of radial integrals, (see formulae (27.2)
- (27.4) there), and our expressions (\ref{eq:b-dd}) and (\ref{eq:b-e}) are
equivalent to the operators they used (except that we have used second
quantization). Jucys and Savukynas~\cite{JS} had also presented matrix
elements of exchange terms for two open shells case in their (27.7)-(27.9).
Their operators for these cases correspond to our operators for
distributions $\alpha \beta \beta \alpha $ and $\beta \alpha \alpha \beta $.
For these distributions there are no vanishing tensorial structures in the
spin-other-orbit interaction operator, so the simplification mentioned above
is no longer possible. Then we directly use a general approach as described
in P2~\cite{GRF}.

\section{Conclusions}

The tensorial form of the spin-other-orbit interaction operator in the
formalism of second quantization is presented (expressions (\ref{eq:m-i}) (%
\ref{eq:b-b}), (\ref{eq:b-c}) and (\ref{eq:b-d})). This tensorial form
allows one to exploit all the advantages of the approach described by
P2~\cite{GRF}:

i) obtaining both diagonal and off-diagonal matrix elements with respect to
the configurations in a unified approach,

ii) using the tables of submatrix elements of
tensorial operators (standard quantities),

iii) applying and making use of  the quasispin formalism for the second quantized,

iv) having recoupling matrices simpler than in other known approaches.

The operator itself generally contains tensorial structures of six different
types: $H_{soo}^{\left( k-1k1,101\right) }$, $H_{soo}^{\left(
k-1k1,011\right) }$, $H_{soo}^{\left( kk1,101\right) }$, $H_{soo}^{\left(
kk1,011\right) }$, $H_{soo}^{\left( k+1k1,101\right) }$ and $H_{soo}^{\left(
k+1k1,011\right) }$ (Section 2). Each type of tensorial structure is
associated with different type of recoupling matrix and with different
matrix elements of standard tensorial quantities. Although the approach
of P2~\cite{GRF} allows one to obtain these
quantities fairly efficiently, still it is expedient to simplify the
tensorial form of a complex operator whenever possible. In the present work
we have succeeded in obtaining simpler expressions, having fewer tensorial
structures ($H_{soo}^{\left( k-1k1,101\right) }$, $H_{soo}^{\left(
k-1k1,011\right) }$, $H_{soo}^{\left( k+1k1,101\right) }$ and $%
H_{soo}^{\left( k+1k1,011\right) }$), for some special distributions
(Section 4) for this particularly complex
spin-other-orbit interaction operator,
This facilitates practical calculations of matrix elements
without restraining the generality, and is one more advantage of approach
P2~\cite{GRF}, complementing those already
mentioned.

\section*{\bf Acknowledgements}

This work is part of co-operative research project funded by National
Science Foundation under grant No. PHY-9501830 and by EURONET PECAM
associated contract ERBCIPDCT 940025.

\end{document}